\documentclass[aps,pre,twocolumn,superscriptaddress]{revtex4-2}

\usepackage[T1]{fontenc}
\usepackage[utf8]{inputenc}
\usepackage{amsmath,amssymb}
\usepackage{siunitx}
\usepackage{graphicx}
\usepackage{xcolor}
\usepackage[normalem]{ulem}
\usepackage{enumitem}
\usepackage{wasysym}
\usepackage{amsthm}
\usepackage{tikz}
\newcommand{\BPT}{\mathrm{BPT}}

\newcommand{\IP}{\mathrm{IP}}
\usetikzlibrary{arrows.meta,positioning}
\definecolor{bluecolor}{rgb}{0,0,1}

\usepackage[colorlinks=true,linkcolor=blue,citecolor=blue,urlcolor=blue]{hyperref}

\begin{document}

\title{Residual Saturation under Pressure-Controlled Drainage}

\author{Fernando Alonso-Marroquin}
\affiliation{QuantumFI, Katoomba NSW 2780, Australia}
\affiliation{CIPR, King Fahd University of Petroleum and Minerals, Dhahran 31261, Kingdom of Saudi Arabia}
\email{fernando@quantumfi.net}
\author{Hans J. Herrmann}
\affiliation{Departamento de Fisica, Universidade Federal do Ceara, 60451-970, Fortaleza, Ceara, Brazil}
\affiliation{PMMH, ESPCI, 7 quai St. Bernard, 75005 Paris, France}
\email{hans@ifb.baug.ethz.ch}

\begin{abstract}
Here, pressure-controlled drainage is formulated as bond percolation with trapping on the pore-network graph, establishing a direct connection between percolation theory and pressure--saturation relations. In two dimensions, the deviation of the residual saturation from its non-vanishing thermodynamic limit obeys a finite-size scaling law with exponent $\delta \approx 0.25$, independent of microscopic details of the lattice. In three dimensions, finite-size corrections decay more rapidly ($\delta \approx 0.75$), while the asymptotic residual saturation remains finite and depends on coordination number. This extends the standard invasion-percolation picture beyond the breakthrough state, where the invading cluster is fractal and the invaded-phase saturation vanishes in the infinite-size limit.
\end{abstract}


\maketitle
\thispagestyle{empty}


Liquid invasion into porous media controls fluid redistribution in subsurface formations and underlies a wide range of natural and engineered transport processes. In the capillary-dominated regime, drainage is governed by pore-scale entry thresholds, intermittent burst events, and the accessibility of the pore space to the advancing interface \cite{Singh2019ARFM,Berg2013PNAS}. Recent experiments and simulations have clarified the roles of Haines jumps, cooperative filling, and front morphology in disordered media \cite{Berg2013PNAS,HoltzmanSegre2015PRL,Schluter2016WRR,Holtzman2020CommPhys,Bultreys2024PNAS}. However, the infinite-size limit of residual saturation under pressure-controlled drainage remains unresolved. 
This is the large-system counterpart of a central field-scale question in waterflooding: whether residual oil saturation remains finite as the invaded region approaches reservoir dimensions, given that primary and secondary recovery together often leave about 60--70\% of the original oil in place \cite{Nagy2025CEOR}.

The standard reference model for capillary invasion is invasion percolation \cite{WilkinsonWillemsen1983}. In that framework, the interface advances throat by throat through the locally most accessible path, producing a ramified invading structure whose breakthrough state is governed by fractal geometry \cite{Wilkinson1986PRA,Moura2017PRL}. If breakthrough occurs once the invading cluster spans the sample, then the invaded-phase saturation scales as
\begin{equation}
S_{I,b}^{\IP}(L)\sim \frac{L^{D_f}}{L^d}
= L^{-(d-D_f)} \to 0
\qquad \text{as} \qquad L\to\infty,
\label{eq:SI_intro_IP}
\end{equation}
where \(S_{I,b}^{\IP}\) is the invaded-phase saturation at breakthrough, \(L\) is the system size, \(d\) is the spatial dimension, and \(D_f<d\) is the fractal dimension of the invading cluster.
Previous studies on slow drainage has reinforced the importance of intermittent bursts and nonlocal filling mechanisms \cite{HoltzmanSegre2015PRL,Moura2017PRL}, but these studies do not address the final state of pressure-controlled drainage, in which the invading phase fills all pore regions accessible at the imposed capillary threshold.

This distinction is closely tied to the difference between geometric availability and hydraulic accessibility in the pressure--saturation relation \cite{Holtzman2020CommPhys}. Under pressure-controlled drainage, the relevant issue is not only whether a throat is open in principle, but whether the corresponding pore remains connected to the inlet while the defending phase still retains a path to the outlet. In pore-network terms, trapping is therefore determined by connectivity \cite{Petrovskyy2021AWR}. This observation motivates a graph-based description in which capillary thresholds define  the pressure-accessible subnetwork, inlet connectivity determines invasion, and outlet disconnection determines trapping.

A second motivation is scale dependence. Recent pore-scale measurements show that drainage-induced perturbations extend beyond a single pore, with direct implications for representative elementary volumes and for the interpretation of macroscopic constitutive laws \cite{Bultreys2024PNAS}. The size dependence of residual saturation is therefore part of the physical problem rather than a purely numerical detail. Here we show that pressure-controlled drainage leads instead to a space-filling invaded state with a finite residual saturation in the infinite-size limit,
\begin{equation}
S_r(L)\to S_\infty,
\qquad
0<S_\infty<1,
\qquad
\text{as}
\qquad
L\to\infty,
\label{eq:Sr_intro_limit}
\end{equation}
where \(S_r\) is the residual saturation of the defending phase and \(S_\infty\) is its thermodynamic-limit value. The condition \(0<S_\infty<1\) implies that, under pressure-controlled drainage, a finite fraction of the pore space is invaded while a finite fraction remains trapped even in the infinite-system limit. By contrast, invasion percolation at breakthrough predicts \(S_\infty=1\), since the invading-phase saturation vanishes as \(L\to\infty\); in that limit essentially all of the defending fluid remains in place and only an infinitesimal fraction is recovered.


\begin{figure*}[t]
    \rotatebox{90}{Percolation with trapping}
    \includegraphics[trim={5cm 7.5cm 4cm 6cm},clip,width=0.24\linewidth]{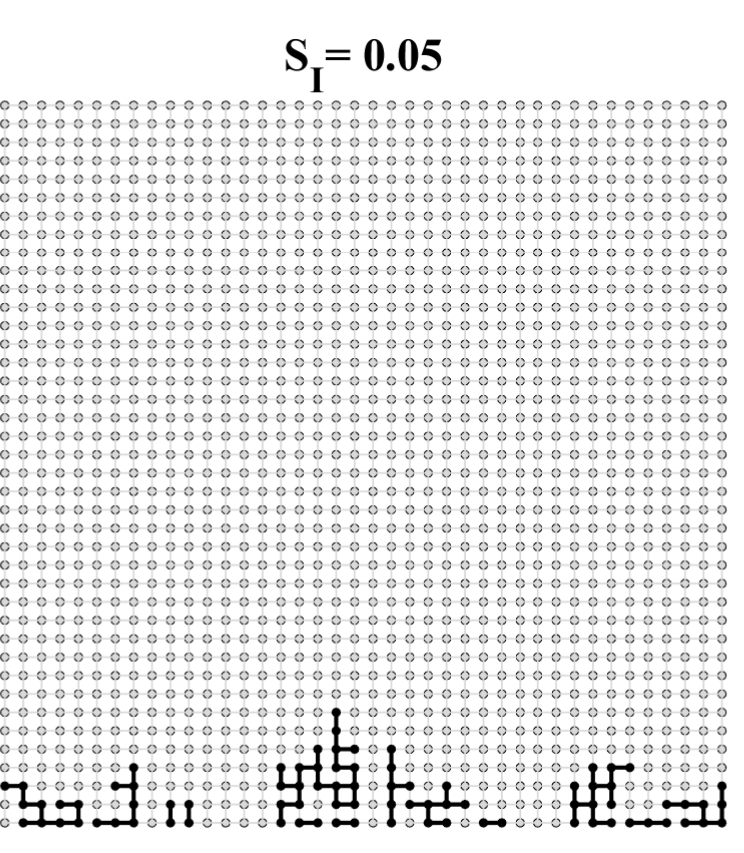}
    \includegraphics[trim={5cm 7.5cm 4cm 6cm},clip,width=0.24\linewidth]{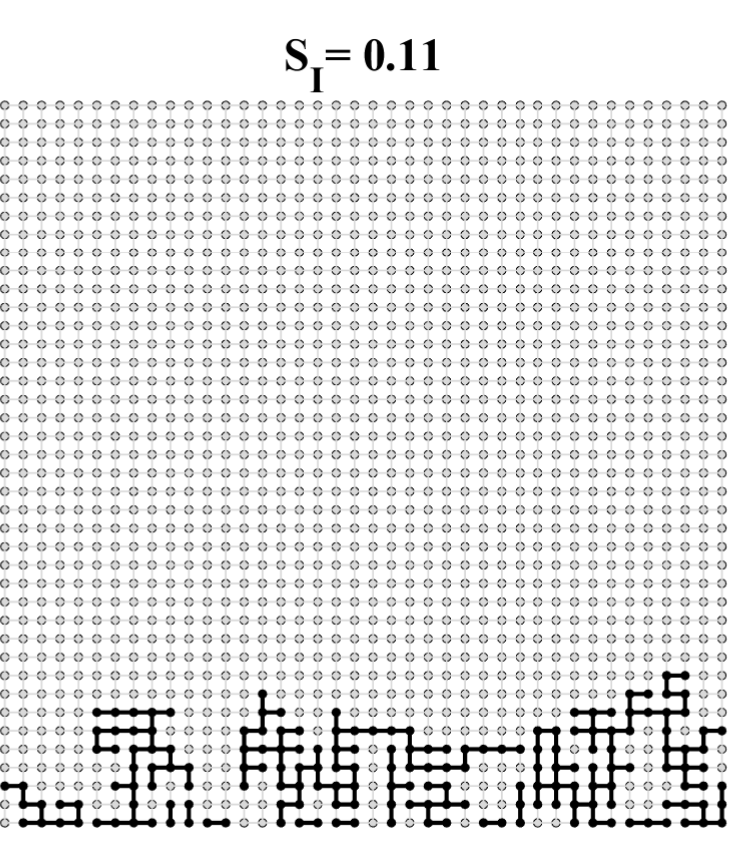}
    \includegraphics[trim={5cm 7.5cm 4cm 6cm},clip,width=0.24\linewidth]{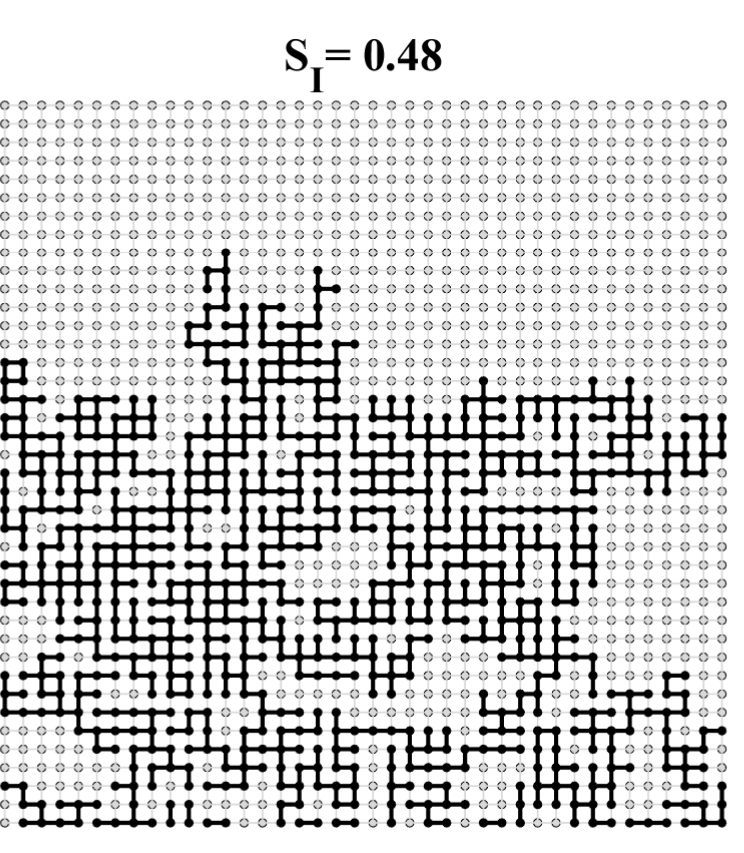}
    \includegraphics[trim={5cm 7.5cm 4cm 6cm},clip,width=0.24\linewidth]{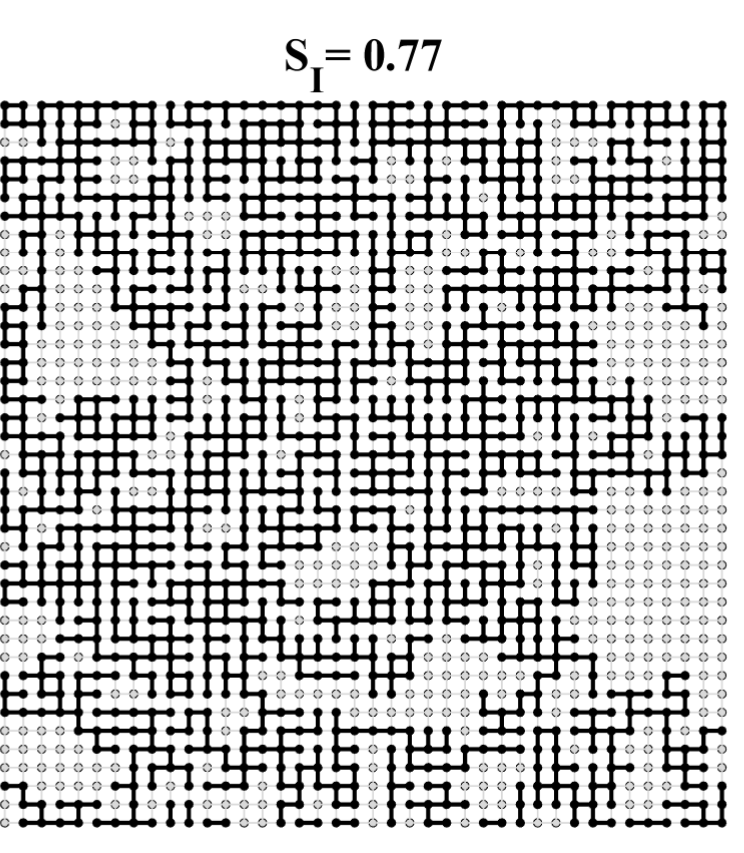}
    \\
    \rotatebox{90}{Invasion percolation}
    \includegraphics[trim={5cm 7.5cm 4cm 6cm},clip,width=0.24\linewidth]{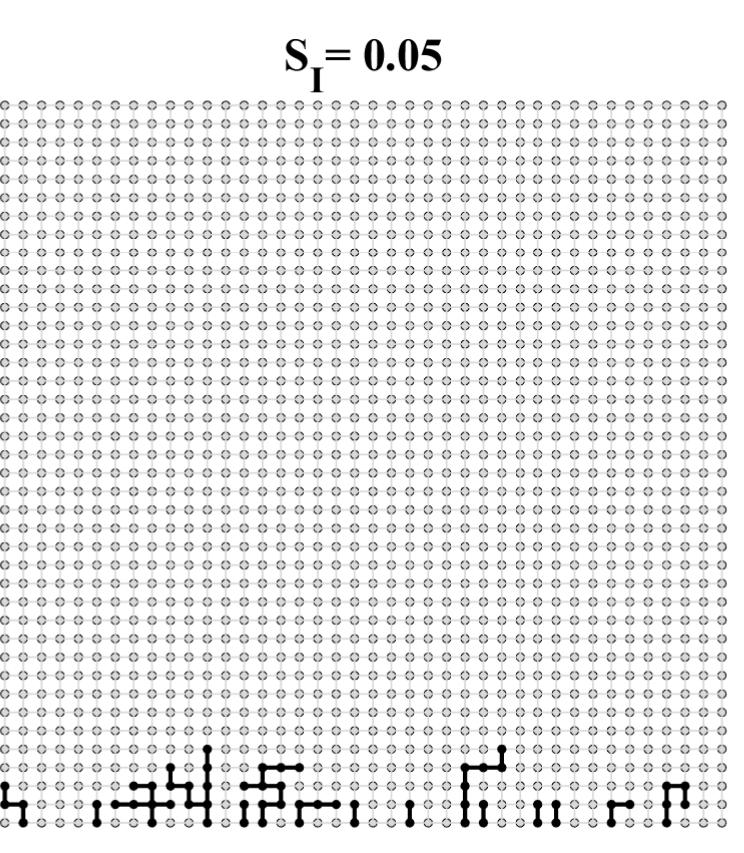}
    \includegraphics[trim={5cm 7.5cm 4cm 6cm},clip,width=0.24\linewidth]{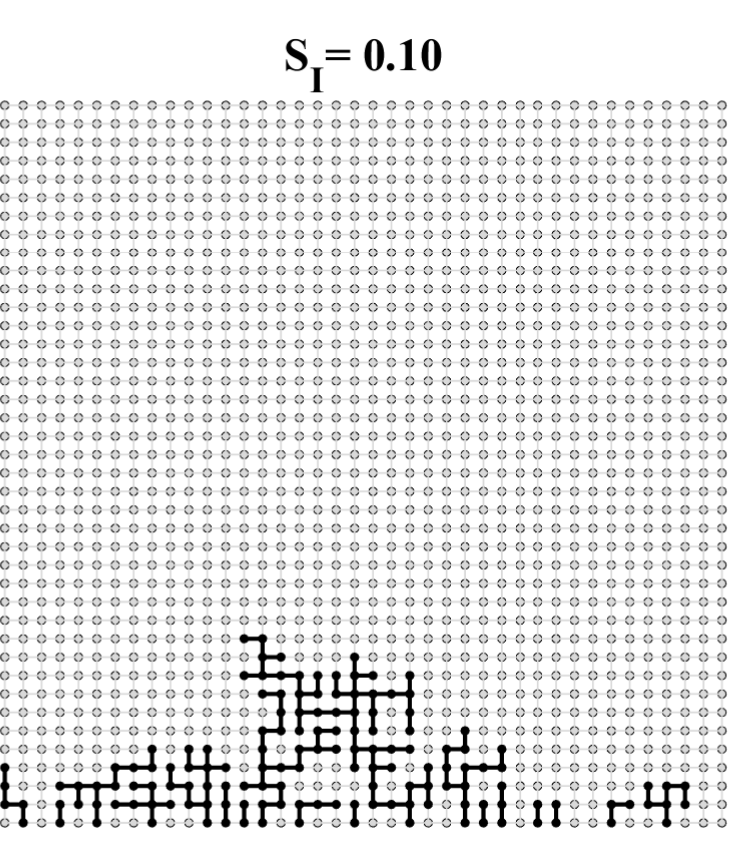}
    \includegraphics[trim={5cm 7.5cm 4cm 6cm},clip,width=0.24\linewidth]{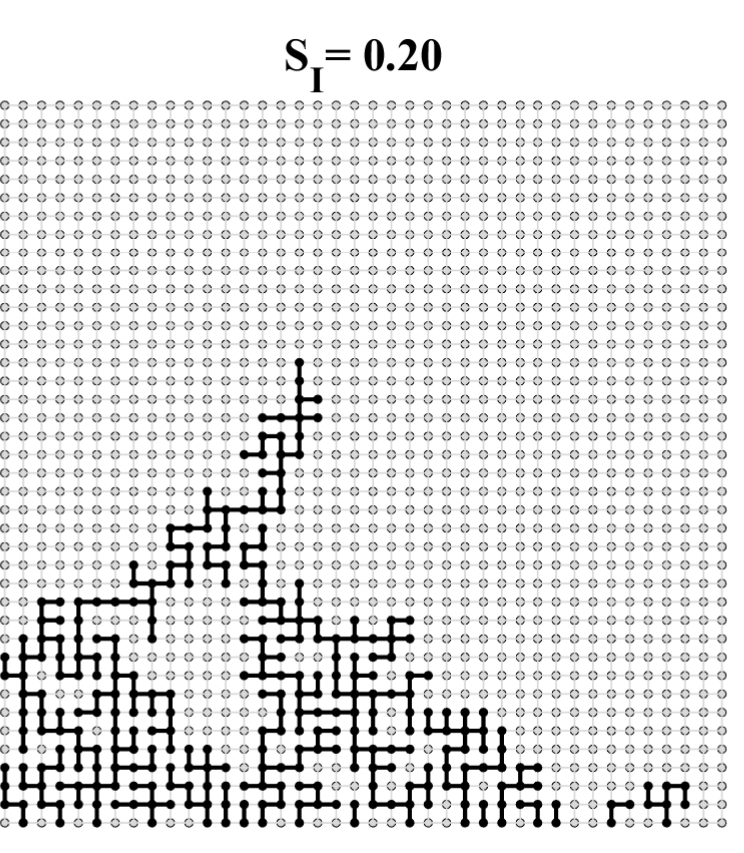}
    \includegraphics[trim={5cm 7.5cm 4cm 6cm},clip,width=0.24\linewidth]{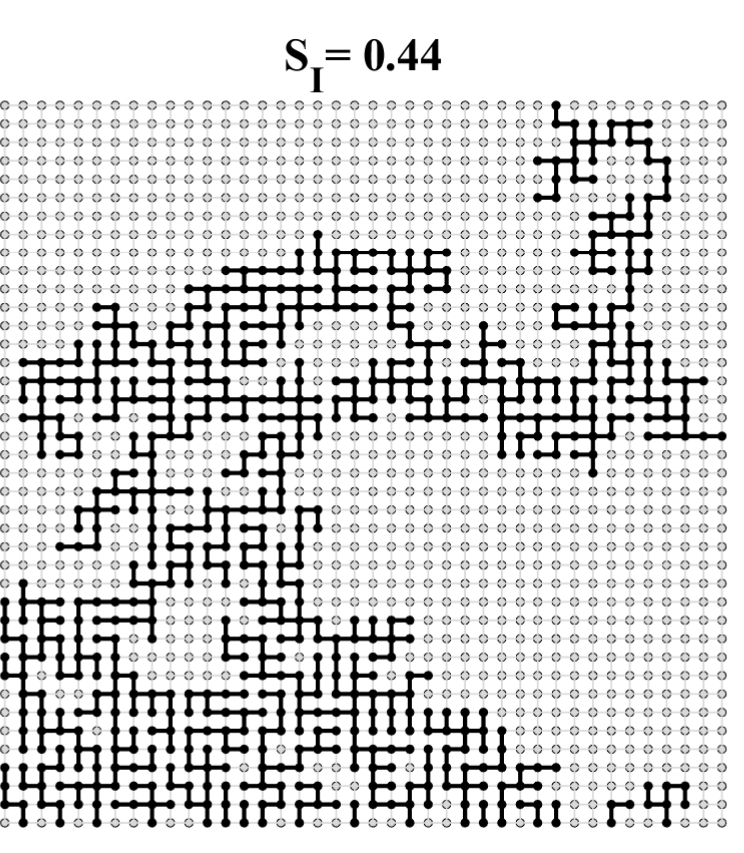}
    \caption{Snapshots of invasion on the same random porous network at different saturation stages for bond percolation with trapping (BPT, top row) and invasion percolation (IP, bottom row).  The bottom and top boundaries define the inlet and outlet nodes, while the left and right boundaries are impermeable no-flow boundaries, i.e., periodic boundary conditions are not used in the horizontal direction. The invaded-phase saturation $S_I$ is the portion of the network reached from the inlet. The final states (last column) show $S_I=77\%$ for BPT, due to trapping of defending-fluid regions disconnected from the outlet, and $S_I=44\%$ for IP, due to early breakthrough and bypassing by the ramified invading cluster. The $p-$increment in BPT is $\Delta p=10^{-3}$.}
    \label{fig:snapshots}
\end{figure*}

\textit{Model.---}
\label{sec:model}
We consider quasi-static drainage in a porous medium represented by an undirected pore-network graph \(g=(V,E)\), where nodes \(V\) represent pore regions and edges \(E\) represent throats connecting adjacent pores. Each throat \(e\in E\) is assigned an entry radius \(r_e\), which determines its capillary threshold. Two boundary sets are distinguished: inlet nodes \(R\subset V\), in contact with the invading phase, and outlet nodes \(O\subset V\), through which the defending phase may escape. Each pore is treated as either invaded or non-invaded, so the saturation is a binary pore occupancy on the graph. Details of the identification of pores and throats from binary images are provided in the Supplemental Material.

The capillary threshold is determined by the Young--Laplace relation,
\begin{equation}
P_c = \frac{2\gamma\cos\theta}{r_c},
\label{eq:YL}
\end{equation}
where \(P_c\) is the capillary pressure, \(\gamma\) is the interfacial tension, \(\theta\) is the drainage contact angle, and \(r_c\) is the capillary radius. A throat \(e\) of radius \(r_e\) is active at pressure \(P_c\) if \(r_e>r_c\). The active graph \(g_p\subseteq g\) is obtained by retaining only the active throats. If \(F(r)\) denotes the cumulative distribution function of throat radii on the graph, the corresponding bond occupancy, defined as the fraction of active throats, is
\begin{equation}
p = 1-F(r_c).
\label{eq:occ}
\end{equation}
Increasing pressure therefore progressively opens the throat network. The graph-based order parameter is the invaded-phase saturation,
\begin{equation}
S_I = \frac{N_{p,I}}{N_p},
\label{eq:SI}
\end{equation}
where \(N_{p,I}\) is the number of invaded pores and \(N_p\) is the total number of pores.
As \(p\) increases, invasion propagates through inlet-connected paths of active throats. A newly activated throat can connect the inlet to a larger active cluster, producing a cascade in which many pores become invaded at nearly the same threshold. Trapping occurs when a portion of the defending phase loses connectivity to the outlet set \(O\); once disconnected, that cluster remains trapped. A pore-level visualization of this cascade and trapping mechanism is provided in the Supplemental Material.

We formulate this process as bond percolation with trapping (BPT). Invasion is determined by inlet connectivity on the active graph, whereas trapping is determined by loss of outlet connectivity of the defending phase in the remaining non-invaded network. In the absence of trapping, every pore connected to the inlet set \(R\) through the active graph is classified as invaded. With trapping, the pore network is partitioned into invaded, defending, and trapped regions. The BPT update is:

\begin{enumerate}[noitemsep]
\item Increase the bond occupancy \(p\), equivalently increasing the capillary pressure.
\item Build the active graph by retaining only the throats active at that pressure.
\item Update the invaded region by adding all pores in the active graph that are not trapped and are connected to the inlet.
\item Update the defending region as the part of the non-invaded network connected to the outlet.
\item Add to the trapped region all pores that are neither invaded nor defending.
\item Repeat until \(p=1\).
\end{enumerate}

The \(p\)-increments in BPT can be defined from an event-based sequence of threshold levels.
Thus, the increment $\Delta p$ is a numerical resolution parameter for the pressure (or occupancy) sweep.  For a fixed porous medium with a continuous distribution of throat thresholds, taking $\Delta p\to 0$ corresponds to resolving the activation of individual throats in order of their entry thresholds. 
In this quasi-continuous limit, a newly activated throat does not necessarily produce invasion; invasion occurs only when that throat connects additional pore space to the inlet-accessible set. 
Near the connectivity transition, a single such connection can trigger an \emph{avalanche} (a cascade) in which many pores become inlet-accessible at essentially the same capillary threshold. 
This update mechanism at fixed pressure avoids Haines-jump behavior discussed in the invasion-percolation literature~\cite{wilkinson1983invasion,dias1986percolation}. 
Numerically, reproducing this limit requires $\Delta p$ to be smaller than the difference between the 
capillary entry pressures of any pair of throats in the system, which does decrease with system size.

The BPT procedure defines the pressure--saturation relation. Because the bond occupancy is linked to capillary pressure through Eqs.~\eqref{eq:YL} and \eqref{eq:occ}, the invaded-phase saturation may be viewed either as a function \(S_I(p)\) or as a function \(S_I(P_c)\). As the active graph grows monotonically with pressure, the defending network shrinks through invasion; outlet connectivity can then be lost, producing trapped regions.
\\

\begin{figure}[t]
    \centering
    \includegraphics[trim={1cm 7cm 2cm 8cm},clip,width=\linewidth]{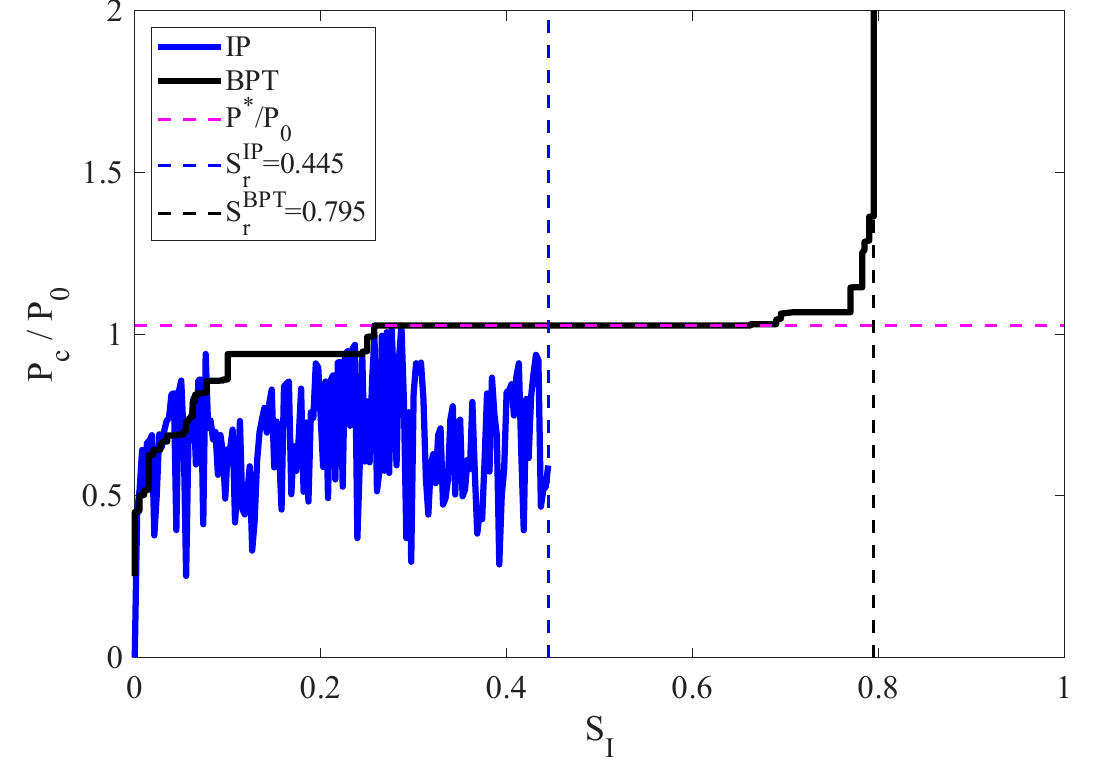}
    \caption{Pressure--saturation curves for the square lattice of Fig.~\ref{fig:snapshots}. black: bond percolation with trapping. Blue: invasion percolation. dashed horizontal line is the entry pressure. The dashed vertical lines indicate the final residual saturation for the trapped percolation model and invasion percolation.The $p-$increment is $\Delta p=10^{-3}$.}
    \label{fig:pressure-saturations}
\end{figure}

\textit{Results.---}
\label{sec:results}

\begin{figure*}[t]
    \centering
    \includegraphics[trim={4cm 7cm 3cm 5cm},clip,width=0.25\linewidth]{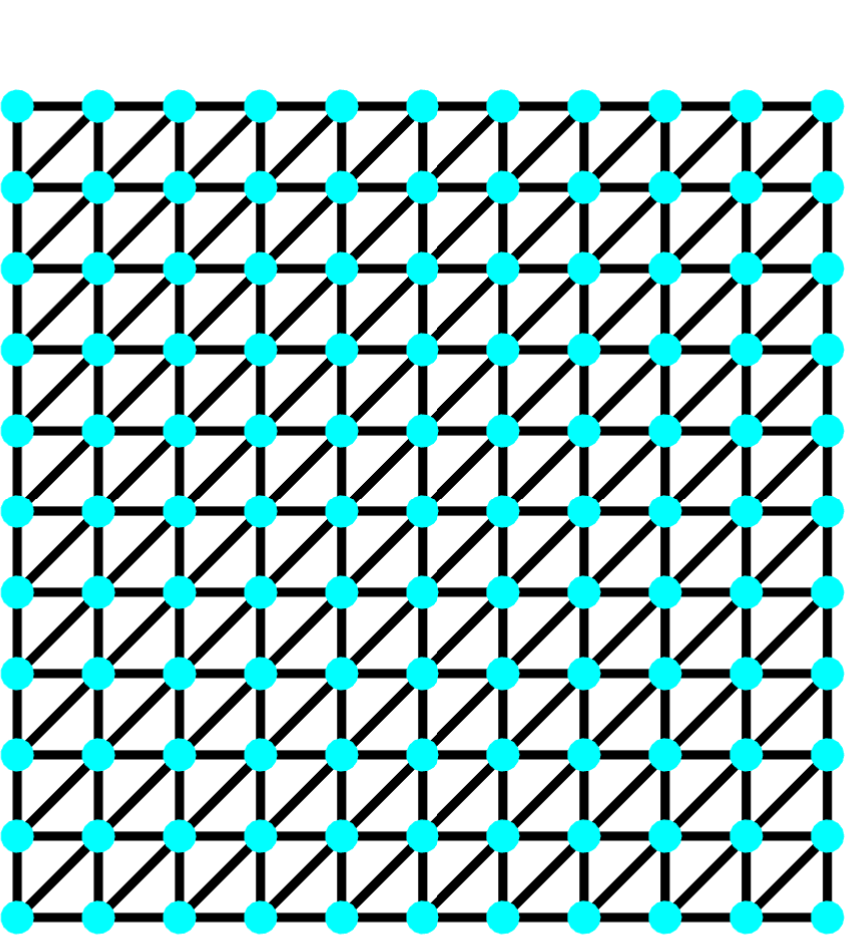}\qquad
    \includegraphics[trim={4cm 7cm 3cm 5cm},clip,width=0.25\linewidth]{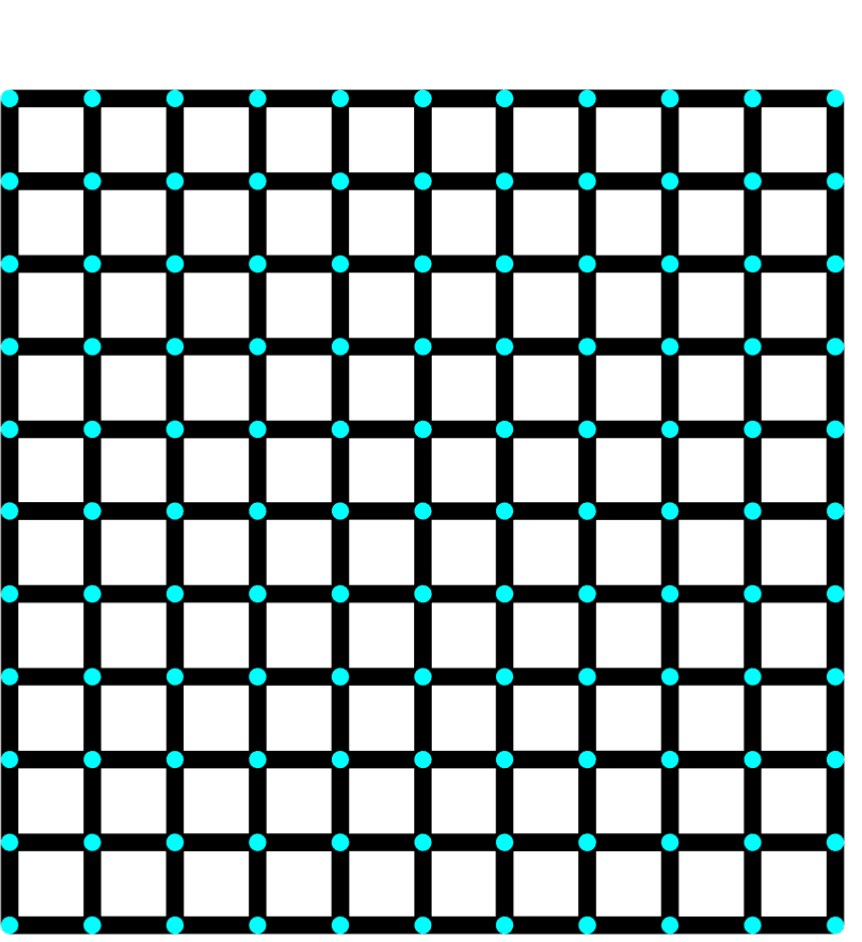}\qquad
    \includegraphics[trim={4cm 7cm 3cm 5cm},clip,width=0.25\linewidth]{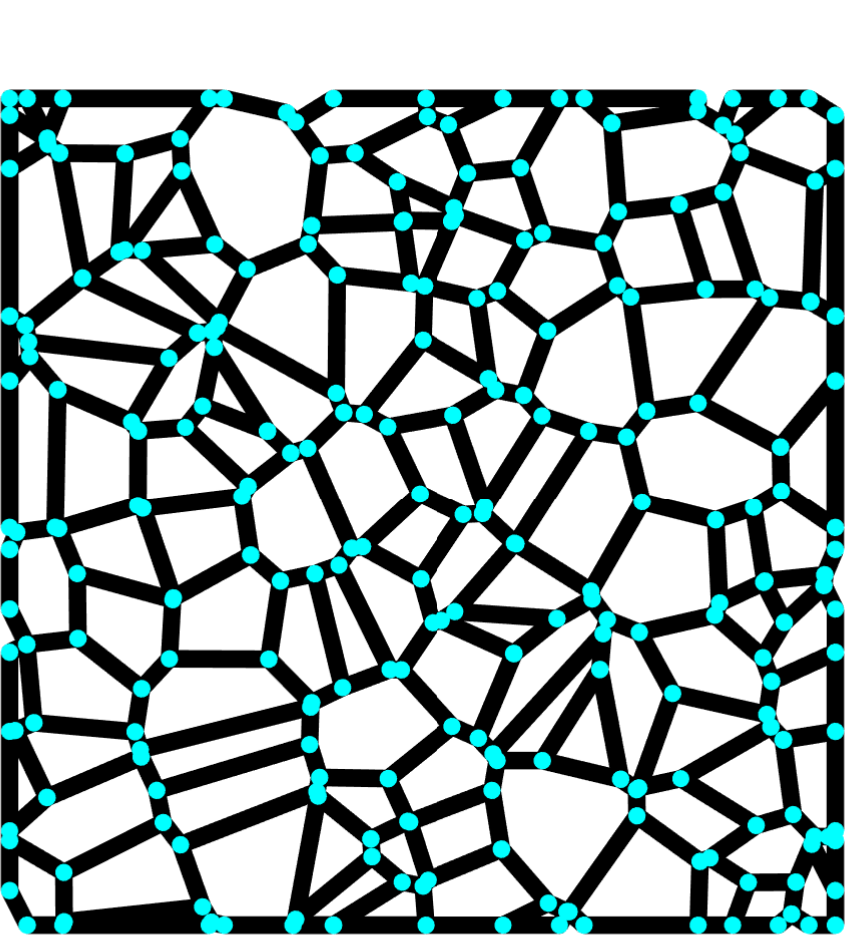}\\
     \includegraphics[trim={0cm 3cm 2cm 4cm},clip,width=0.4\linewidth]{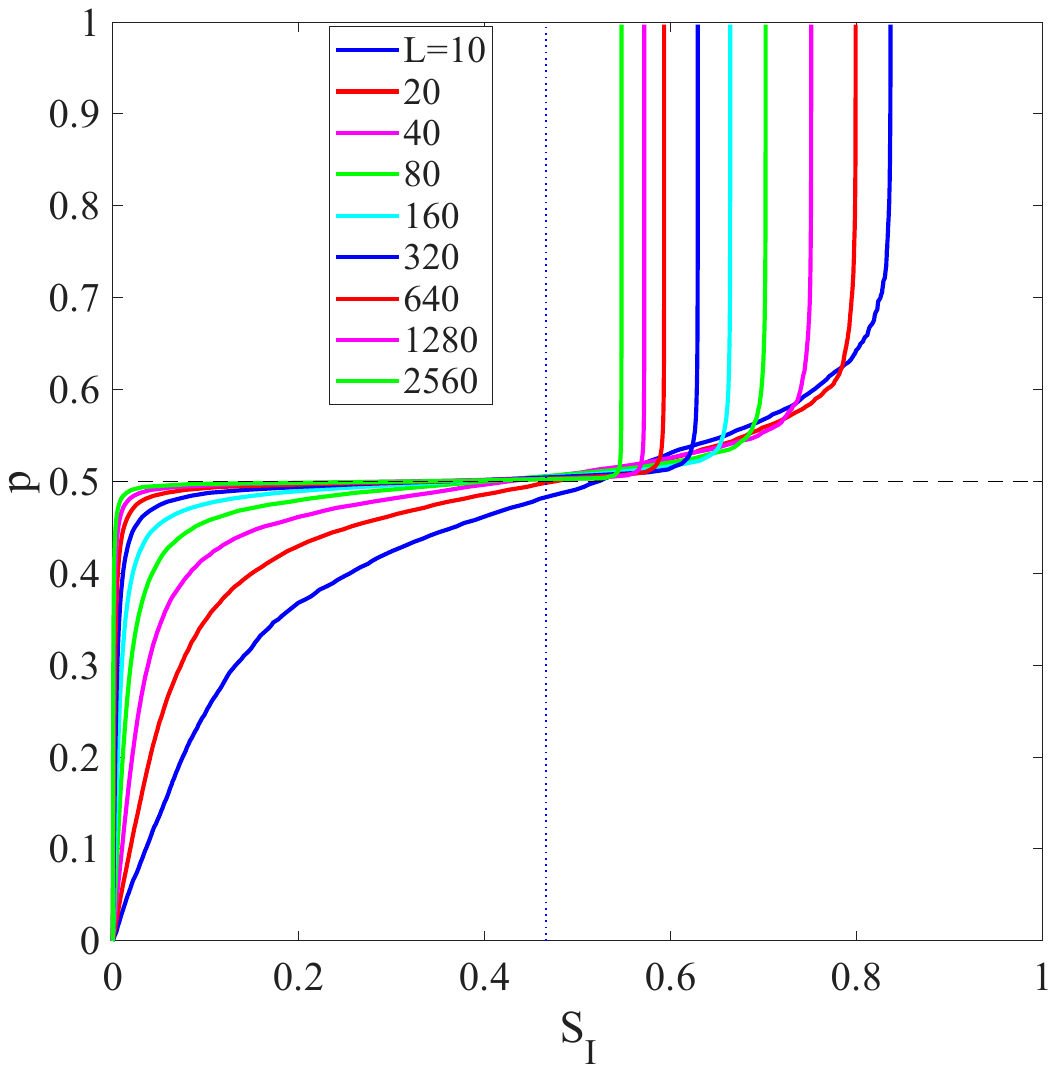}
    \includegraphics[trim={0.5cm 6cm 1cm 7cm},clip,width=0.5\linewidth]{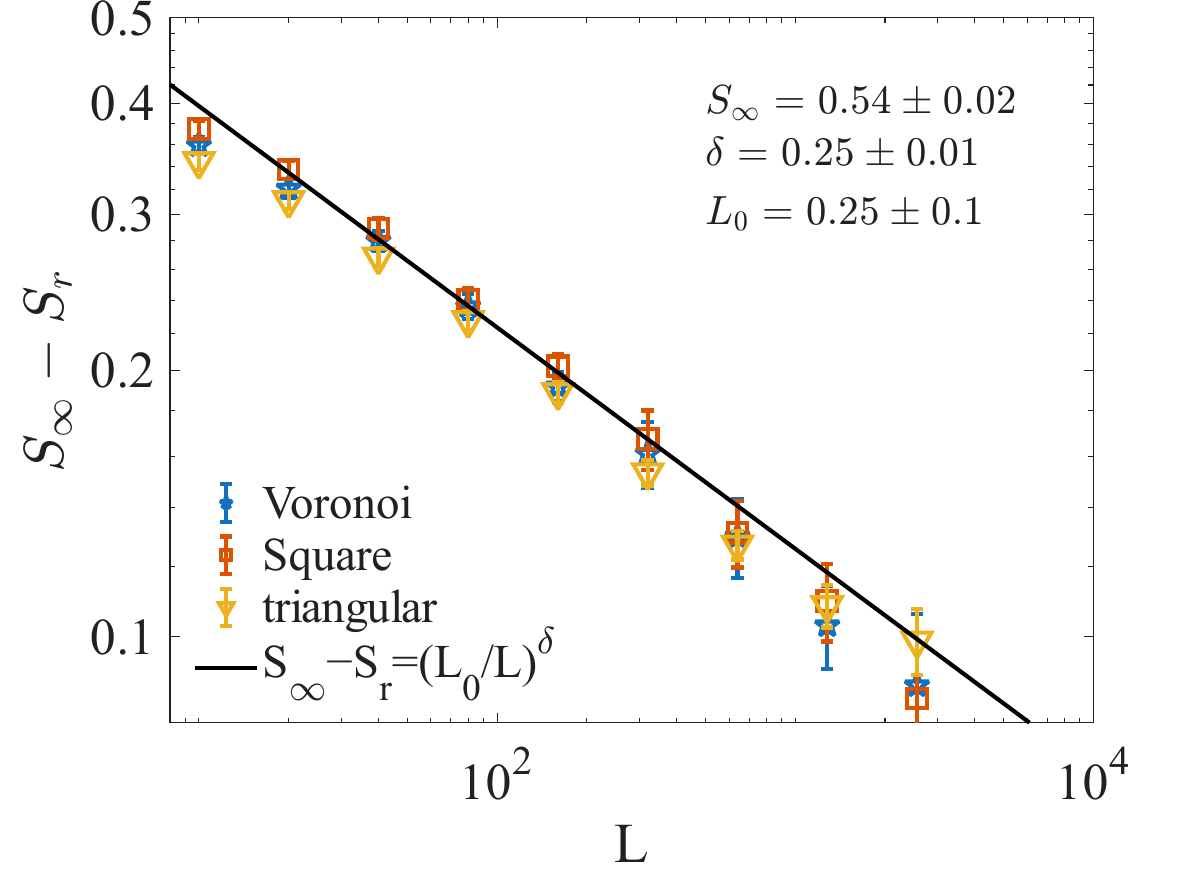}
   
    \caption{Top: lattices used in the 2D analysis, from left to right: triangular, square, and Voronoi. 
    Bottom left: bond occupancy versus invaded-phase saturation for square lattices of increasing size. The dashed lines indicate the classical bond-percolation threshold. The dotted line marks the infinite-system value $S_{I,\infty}(1)=1-S_{\infty}\approx 0.46$.   
    Bottom right: residual saturation versus lattice size for triangular (triangles), square (squares), and Voronoi (stars) lattices. Here $S_r(L)$ denotes the residual saturation for a system of size $L$, and $S_\infty$ is its infinite-size limit value. The solid line is the fit of Eq.~\ref{eq:residual}. The $p-$increment in BPT is $\Delta p=10^{-3}$.}
    \label{fig:lattices}
\end{figure*}

We now address a central question of quasi-static drainage: \textit{can pressure-controlled invasion produce a nonvanishing residual saturation in the infinite-system limit, and if so, what mechanism is responsible beyond the standard invasion-percolation picture?} To answer this question, we compare invasion percolation (IP) with bond percolation with trapping (BPT).

Figure~\ref{fig:snapshots} shows snapshots of the IP and BPT simulations on a square lattice. The lattice has a side length of 0.1 mm and consists of $40 \times 40$ cells. The throat radii follow a lognormal distribution with a median $r_0 = 0.01$ mm and $\sigma = 0.5$. The bottom boundary is connected to the injection zone and the top to the evacuation zone. From Eqs.~\eqref{eq:YL} and \eqref{eq:occ}, the occupancy $p$ is the fraction of active throats,
\begin{equation}
p = 1 - F\left( \frac{2\gamma \cos\theta}{P_c} \right),
\label{eq:Pc-p}
\end{equation}
where $F(r)$ is the cumulative distribution of throat radii. This relation links the applied pressure to the accessible pore space, so that increasing $P_c$ progressively opens more of the network.

The first row of Fig.~\ref{fig:snapshots} shows BPT. As in bond percolation, invasion is determined by connectivity through the subset of active throats defined by the occupancy $p$. Through Eq.~\eqref{eq:Pc-p}, this connectivity problem is here driven by capillary pressure.  In the first row of Fig.~\ref{fig:snapshots}, a significant portion of the network remains occupied at maximum pressure, yielding a final invaded-phase saturation of $S_{I,f}^{\BPT} = 77\%$ and residual saturation $S_r^{\BPT} = 23\%$. This outcome reflects the irreversible nature of disconnection in incompressible fluids.

In contrast, IP progresses via the selective invasion of the largest accessible throat at each step, forming a fractal, tortuous cluster. The entry pressure at each step is defined by $P_t = 2\gamma \cos\theta / r$, and the corresponding saturation is computed using Eq.~\eqref{eq:SI}. As shown in the second row of Fig.~\ref{fig:snapshots}, breakthrough occurs when a spanning path emerges, here at $S_{I,b}^{\IP} = 44\%$, leaving a residual saturation of $S_r^{\IP} = 56\%$. IP therefore produces a larger uninvaded-phase saturation at breakthrough because the invasion process favors early spanning and bypass of large regions, whereas BPT describes the final trapped state under pressure-controlled invasion.

The corresponding pressure--saturation curves are shown in Fig.~\ref{fig:pressure-saturations}. Invasion percolation yields a jagged sequence of entry events, reflecting the burst-like advance of the interface. Pressure-controlled drainage yields a much smoother response because invasion is determined by accessibility over the whole active network at each pressure level. The essential distinction is therefore not merely that the curves have different shapes, but that they terminate at different physical endpoints: breakthrough in invasion percolation and full pressure access in the trapped percolation model.

\paragraph{Finite-size effects}
\label{sec:finite}

We now examine how the invasion curve and the final trapped fraction depend on system size. Finite-size effects are intrinsic to percolation processes \cite{hunt2017flow,stauffer2018introduction}, and are especially relevant in porous media because both experiments and simulations are necessarily performed on finite samples. Establishing how the residual saturation approaches its large-system limit is therefore essential for connecting pore-scale invasion to macroscopic behavior.

The same finite-size signatures that sharpen the bond-percolation transition are observed here in BPT. As the system size increases, the transition in the $S$--$p$ relation becomes sharper, while sample-to-sample fluctuations near the transition remain strong, requiring ensemble averaging. We therefore performed simulations on three classes of 2D networks---square, triangular, and Voronoi, shown in Fig.~\ref{fig:lattices}---which have well-defined bond-percolation thresholds \cite{kesten1980critical,sykes1963some,becker2009percolation} and belong to the same universality class. The linear size is defined as $L=\sqrt{N_p}$, where $N_p$ is the number of pores. We varied the system size as $L=10\times 2^n$, with $n=0,1,\dots,8$, and for each case averaged over $960$ realizations with lognormally distributed throat radii.

Figure~\ref{fig:lattices} shows that, as $L$ grows, the transition becomes sharper but the residual saturation converges only slowly. The lower-right panel shows that, for all three lattices, the finite-size correction is well described by
\begin{equation}
S^{\BPT}_r(L) \approx S_\infty - \left(\frac{L_0}{L}\right)^\delta,
\label{eq:residual}
\end{equation}
with $\delta = 0.25 \pm 0.01$, $L_0 = 0.25 \pm 0.1$, and $S_\infty = 0.53 \pm 0.04$. The collapse of the three lattice families onto the same exponent supports a universal finite-size law for pressure-controlled trapping in two dimensions.

This scaling should not be confused with the fractal scaling of invasion percolation at breakthrough. The exponent $\delta$ in Eq.~\eqref{eq:residual} is a finite-size correction exponent for the final trapped state of BPT at $p=1$; it is neither a critical exponent nor the fractal dimension of a percolation cluster. In particular, the final invaded set in BPT is not fractal. From Eq.~\eqref{eq:residual}, its mass satisfies
\begin{equation}
M_I^{\BPT}(L)=\bigl(1-S_r^{\BPT}(L)\bigr)L^d
\approx (1-S_\infty)L^d + L_0^{\delta}L^{d-\delta},
\label{eq:mass_bpt}
\end{equation}
so the dominant scaling is space-filling, with only subleading corrections.

For comparison, the invasion-percolation cluster at breakthrough is fractal in two dimensions, with numerical estimates $D_f^{\mathrm{IP}}\approx 1.89$ \cite{wilkinson1983invasion}. If breakthrough occurs once the invaded cluster spans a distance of order $L$, then
\begin{equation}
S_{I,b}^{\mathrm{IP}}(L)\sim \frac{L^{D_f^{\mathrm{IP}}}}{L^d}
= \left(\frac{L_0}{L}\right)^{\alpha},
\label{eq:SI_ip_scaling}
\end{equation}
where $\alpha=d-D_f^{\mathrm{IP}}$, so that for $d=2$ one has $\alpha\approx 0.11$. Thus $S_{I,b}^{\mathrm{IP}}(L)\to 0$ at breakthrough, whereas in BPT the final residual saturation approaches a nonzero limit, $S_r^{\BPT}(L)\to S_\infty$, where $0<S_\infty<1$. The key distinction is therefore not the numerical value of the exponent alone, but the asymptotic state itself: invasion percolation yields a fractal cluster, whereas pressure-controlled BPT yields a space-filling invaded region with a finite trapped fraction in the infinite-size limit.

The residual saturation approaches its thermodynamic (infinite-size) limit very slowly. Writing
\begin{equation}
\Delta S^{\BPT}_r(L)\equiv S_\infty-S^{\BPT}_r(L)\approx \left(\frac{L_0}{L}\right)^{\delta},
\label{eq:deltaS_def}
\end{equation}
and using the 2D result $\delta\simeq 0.25$, one obtains $\Delta S_r\sim L^{-0.25}$. As a consequence, even a modest reduction of the finite-size bias requires a large increase in system size. For example, reducing $\Delta S_r$ by a factor of two requires
\begin{equation}
\frac{\Delta S^{\BPT}_r(L_2)}{\Delta S^{\BPT}_r(L_1)}=\frac{1}{2}
\quad\Rightarrow\quad
\frac{L_2}{L_1}=2^{1/\delta}\approx 16
\qquad (d=2).
\label{eq:rev_factor_2d}
\end{equation}
Thus, in 2D, representative-element requirements become severe: suppressing finite-size effects demands domains whose linear extent may approach the continuum grid scale itself.

\paragraph{Dimensionality effects}

\begin{figure}[t]
  \centering
  \includegraphics[trim={1cm 7cm 1cm 8cm},clip,width=\linewidth]{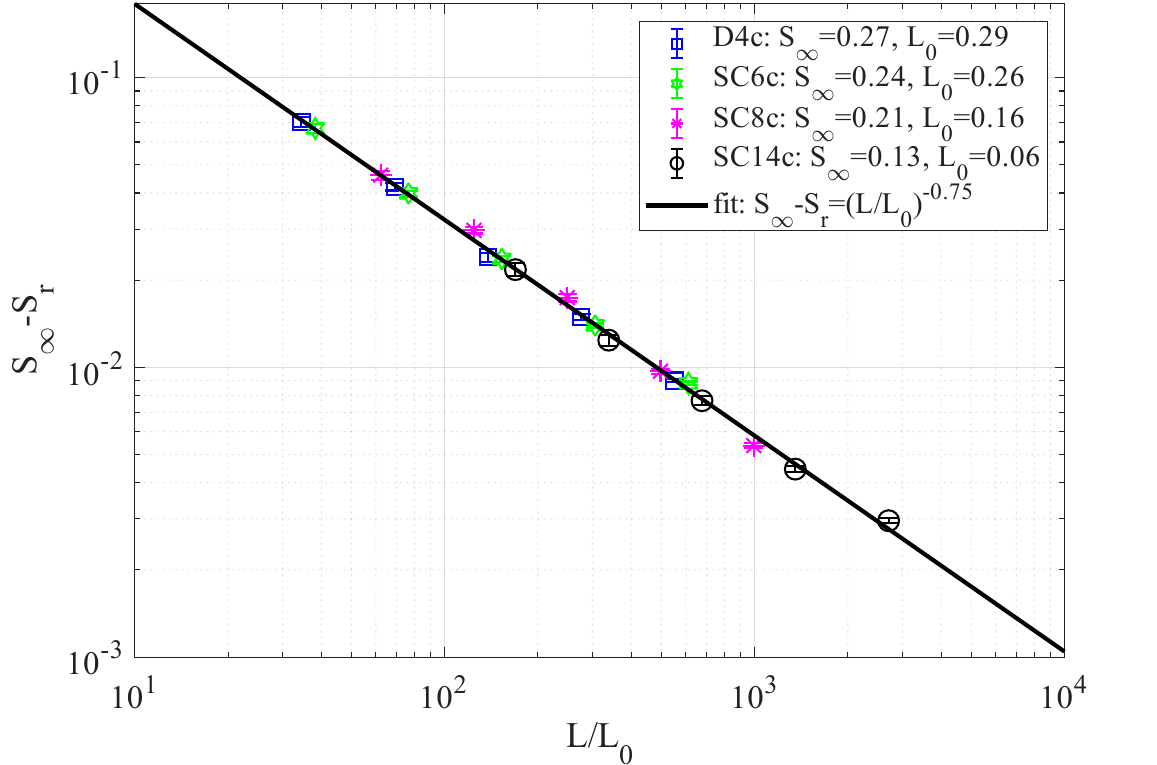}
  \caption{Residual-saturation scaling in 3D pore-network lattices: $S_{\infty}-S_r$ versus normalized system size $L/L_0$ on log--log axes. Results are shown for the four lattices D4c, SC6c, SC8c, and SC14c, where the suffix ``c'' denotes periodic boundary conditions in the transverse directions ($x$ and $y$), while the displacement direction is non-periodic. Symbols show ensemble means over $96$ realizations at each $L=10,20,40,80,160$, with error bars indicating the standard error. The solid line indicates the best-fit power law $(L_0/L)^{\delta}$. The $p-$increment in BPT is $\Delta p=10^{-3}$.}
  \label{fig:residual_saturation_3D}
\end{figure}

Two-dimensional networks are directly relevant to planar micromodels, but natural porous materials are intrinsically three-dimensional. We performed BPT simulations on four three-dimensional regular lattices: the diamond lattice (D4) and simple-cubic lattices with coordination numbers 6 (SC6), 8 (SC8), and 14 (SC14). The SC14 lattice is constructed as the superposition of SC6 and SC8, yielding enhanced connectivity. The results are shown in Fig.~\ref{fig:residual_saturation_3D}.

In 3D the same finite-size form is observed, but with a much larger exponent,
\begin{equation}
\Delta S^{\BPT}_r(L)\approx \left(\frac{L_0}{L}\right)^{0.75}
\qquad (d=3),
\label{eq:deltaS_3d}
\end{equation}
so that finite-size corrections decay much faster than in 2D. A factor-of-two reduction in $\Delta S^{\BPT}_r$ then requires only
\begin{equation}
\frac{L_2}{L_1}=2^{1/0.75}\approx 2.5.
\label{eq:rev_factor_3d}
\end{equation}
The difference between Eqs.~\eqref{eq:rev_factor_2d} and \eqref{eq:rev_factor_3d} shows that dimensionality has a strong practical effect on the representative volume required to approximate the infinite-size limit.

Figure~\ref{fig:residual_saturation_3D} also shows that the limiting residual saturation decreases as the coordination number increases, while remaining finite for all lattices considered. Greater connectivity provides more alternative invasion pathways and therefore reduces, but does not eliminate, the trapped fraction. Dimensionality thus plays a dual role: it controls the rate at which finite-size effects decay, and in three dimensions it makes the asymptotic residual saturation sensitive to coordination number. The overall conclusion is that representative-volume constraints are much less restrictive in 3D than in 2D, while connectivity-controlled trapped fraction approaches a nonzero limit as the system size tends to infinity.

The present results address the asymptotic value of residual saturation in pressure-controlled drainage. By formulating the process as bond percolation with trapping on the pore graph, we show that the residual saturation approaches a nonzero limit as the system size increases. The event-based version of the BPT algorithm complements IP by applying the same ranked capillary thresholds to pressure-controlled activation of the active graph, followed by connectivity and trapping updates. This produces a space-filling invaded region coexisting with a finite trapped fraction in the infinite-size limit.

In two dimensions, the finite-size correction is slow and robust across different lattice families, with exponent $\delta \approx 0.25$. In three dimensions, convergence is much faster, with $\delta \approx 0.75$, and the limiting residual saturation decreases with coordination number while remaining finite. These results identify connectivity and trapping as the factors determining the residual saturation under pressure-controlled drainage. Pore-shape effects, including corner and corner--bridge flow \cite{Lan2024WRR}, remain important for quantitative comparison with experiments, but they modify rather than replace the relation established here.

\begin{acknowledgments}
F.A.-M. acknowledges the Massachusetts Institute of Technology (MIT), where part of this work was developed, and thanks MIT for its hospitality and stimulating research environment.
\end{acknowledgments}

\bibliographystyle{apsrev4-2}
\bibliography{main,percolation}

\end{document}